\documentclass{PoS}
\usepackage{epsf}
\def\pp{p-p~}
\def\sr17{$\sqrt{s}$~=~17~GeV~}
\newcommand{\pbar}{$\rm\overline{p}$}

\newcommand{\s}{$\sqrt{s}$}
\newcommand{\be}{\begin{equation}}
\newcommand{\ee}{\end{equation}}

\def\muB{$\mu_B$}
\def\muS{$\mu_S$}
\title{Particle Production in p-p and Heavy Ion Collisions at Ultrarelativistic Energies.}
\ShortTitle{Antimatter}
\author{\speaker{J.~Cleymans}$^a$, S.~Kabana$^b$, I.~Kraus$^c$, 
H.~Oeschler$^{d,e}$, K.~Redlich$^f$,  N.~Sharma$^{e,g}$\\
\llap{$^a$}UCT-CERN Research Centre and Department  of  Physics,\\ University of Cape Town, Rondebosch 7701, South Africa\\
\llap{$^b$}SUBATECH, 4 rue Alfred Kastler, F-44307 Nantes,
France\\
\llap{$^c$} GSI, Planckstrasse 1, D-64291 Darmstadt, Germany\\
\llap{$^d$} Institut f\"ur Kernphysik, Darmstadt University of
Technology,\\ D-64289 Darmstadt, Germany\\
\llap{$^e$}
European Organization for Nuclear Research (CERN),\\
Geneva, Switzerland\\
\llap{$^f$}{Institute of Theoretical Physics, University of Wroc\l aw, Pl-45204 Wroc\l aw, Poland}\\
\llap{$^g$} Department of Physics, Panjab University, Chandigarh, India}
\abstract{
Recent results related to the chemical  equilibration of hadrons in 
the final state of p-p and  heavy ion 
collisions are reviewed.  }
\FullConference{Kruger2010:  Workshop on Discovery Physics at the LHC,\\
        December 5 - 10 2010\\
         Kruger National Park, South Africa}

\begin{document}
%
\section{Introduction}
After  analysing particle multiplicities in heavy-ion collisions for two decades
a remarkably simple picture has emerged for the chemical freeze-out 
parameters~\cite{stachel,becattini,comparison}. 
\begin{figure}[tbh]
\begin{center}
\includegraphics[width=7.5cm]{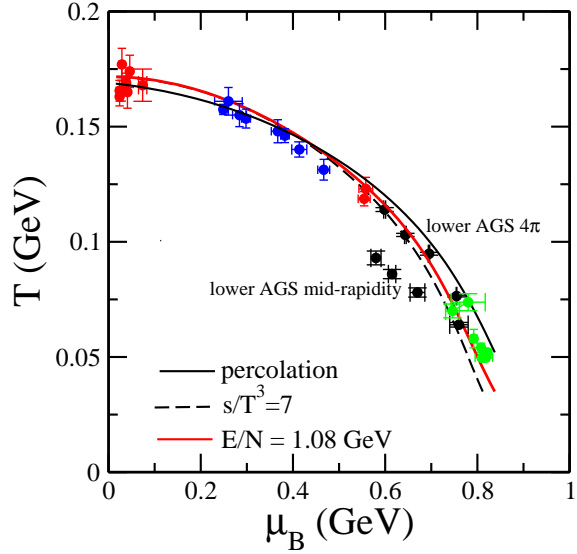}
\label{eovern_2009}
\caption{Values of the freeze-out parameters obtained at beam energies ranging from 1 GeV to 200 GeV}
\end{center}
\end{figure}
Despite much initial skepticism, the thermal model
has emerged as a  reliable guide for particle multiplicities 
in heavy ion collisions at all collision  energies.
Some of the results, including  analyses from~\cite{davis,jun,hades,fopi},
are summarised in Fig.~\ref{eovern_2009}.
 Most of the points in 
Fig.~\ref{eovern_2009} 
(except obviously the ones at RHIC) refer to integrated ($4\pi$)
yields. A clear discrepancy exists in the lower AGS beam energy region
between the  chemical parameters extracted from (published) mid-rapidity 
yields and those extracted using  estimates of the $4\pi$ yields. 
The latter tend to give higher 
values for the chemical freeze-out temperature. This will have to be 
resolved by future experiments at e.g. NICA and FAIR.
When the temperature and baryon chemical potential 
are translated to net baryon and energy densities, 
a different, but equivalent, picture emerges shown in Fig.~\ref{randrup}.
This clearly shows the importance in going to the beam energy region
of around 8 - 12 GeV as this corresponds to the highest freeze-out 
baryonic density 
and to a rapid change in thermodynamic 
parameters~\cite{randrup1,randrup2}.
\begin{figure}
\begin{center}
\includegraphics[width=7.5cm]{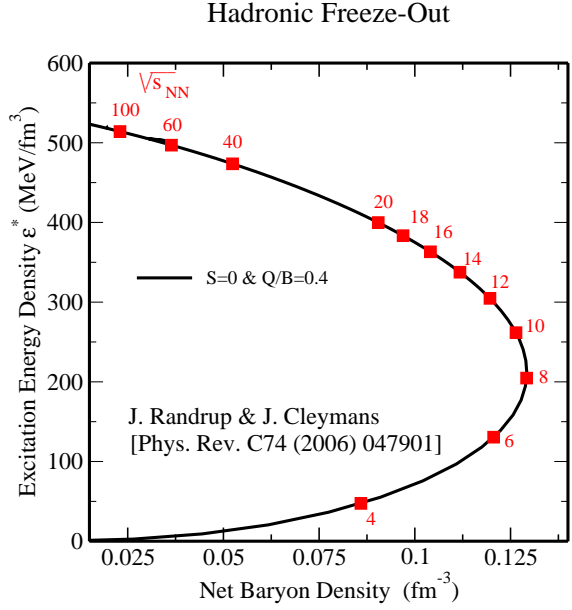}
\caption{The hadronic freeze-out line in the
$\rho_B-\epsilon^*$ 
phase plane as obtained from the values of $\mu_B$ and $T$
 that have been extracted from the experimental data in~\cite{comparison}.
The calculation employs values of $\mu_Q$ and $\mu_S$ 
that ensure $\langle S\rangle=0$ and $\langle Q\rangle=0.4\langle B\rangle$
for each value of $\mu_B$.  
Also indicated are the beam energies (in GeV/N)
for which the particular freeze-out 
conditions are expected at either RHIC or FAIR or NICA. 
}
\label{randrup}
\end{center}
\end{figure}

The dependence of $\mu_B$ on the invariant beam energy, $\sqrt{s_{NN}}$, can be 
parameterized as~\cite{comparison}
$$
\mu_B(\sqrt{s_{NN}}) = \frac{1.308~\mathrm{GeV}}{1 + 0.273~{\mathrm{GeV}}^{-1}\sqrt{s_{NN}}}.
$$
Similar dependences
have been obtained by other groups~\cite{stachel,becattini}. 
and are  consistent with the above.  
This predicts that  at the  LHC $\mu_B\approx 1$~MeV.
\begin{figure}
\begin{center}
\includegraphics[width=7.5cm]{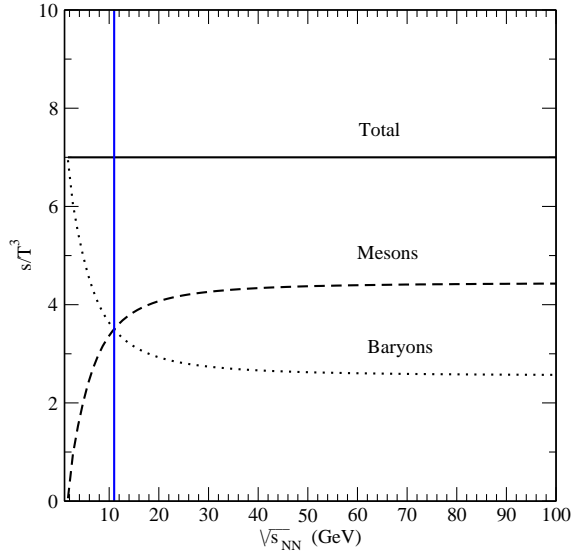}
\caption{Values of entropy density divided by $T^3$  following the 
chemical freeze-out values~\cite{transition}.}
\label{st3}
\end{center}
\end{figure}

To analyze the changes around 10 GeV use can
be made of the 
entropy density, $s$, divided by $T^3$ which has been shown to reproduce the 
freeze-out curve~\cite{comparison} very well. This allows for a separation into 
baryonic and mesonic components, shown in Fig.~\ref{st3}, it can be 
seen that
mesons  dominate the 
chemical freeze-out from about $\sqrt{s_{NN}}\approx$ 10 GeV onwards.
\section{Antimatter Production}
One of the striking features of particle production at high energies
 is the near equal abundance of matter and
antimatter in the central rapidity region~\cite{star:2009,Aamodt:2010d}.
 As is well known a similar symmetry existed in the initial stage of the
universe and it still remains a
mystery as to how this got lost in the evolution of the
universe reaching a stage with no visible
amounts of antimatter being present.\\
Closely related to this matter/antimatter symmetry is the production
of light antinuclei, hypernuclei and antihypernuclei at high energies.
Since the first observation of hypernuclei 
in 1952
\cite{first_hypernucleus} there has been a steady interest in
searching for new hypernuclei and exploring the hyperon-nucleon
interaction which is relevant (see e.g.~\cite{hahn,stoecker}) for nuclear physics.
Hypernuclei decay with lifetime which depends on the strength of
the hyperon-nucleon interaction. While several hypernuclei have
been discovered since the first observations in 1952, no antihypernucleus has ever been
observed until the recent discovery of the antihypertriton in
Au+Au collisions at $\sqrt{s_{NN}}$ = 200 GeV by the STAR collaboration at
RHIC \cite{Abelev:2010}.
The yield of (anti)hypernuclei  measured  by STAR is very large, in particular they seem
to be produced with a  similar yield as other (anti)nuclei, in particular (anti)helium-3.
This  abundance is much higher than measured for hypernuclei and nuclei at 
lower energies~\cite{shuryak}.
It is of interest to understand the nature of this enhancement, and
for this the mechanism of production of (anti)hypernuclei should 
be investigated.

The analysis of particle production assessing the degree of 
thermalization of the particle source
 has been undertaken for many decades
\cite{Fermi,Pomeranchuk,Heisenberg,Landau,Hagedorn}. 
It has been found that the thermalization assumption 
applies  successfully to hadrons
produced in a large number of particle and nuclear reactions 
at different energies
\cite{becattini1,review,energy}.
This fact allows us to estimate thermal parameters characterizing the particle 
source for each colliding system,   relevant for the understanding 
of the thermal properties of dense and hot matter, and in particular
for studies of QCD phase transitions. 
In this paper,
 using the parametrizations of thermal parameters estimated by the model 
THERMUS~\cite{thermus1,thermus2} 
 that were shown to best fit the existing data from particle and nuclear collisions at several energies,
 we make thermal model estimates of (anti)hypernuclei
 that can be directly compared to 
the recently measured unexpected high (anti)hypernuclei yields at RHIC
as well as predictions of (anti)matter and (anti)hypernuclei production at 
the Large Hadron Collider (LHC).  A similar analysis, not including 
\pp  results,  has been presented recently in~\cite{andronic-heavy} 
where it was shown that  ratios of hypernuclei to nuclei show an energy
dependence similar to the $K^+/\pi^+$ one with a clear maximum at lower energies.
In this paper we 
study quantitatively how the matter/antimatter symmetry is reached as the
beam energy is increased. We also  estimate ratios of hypernuclei and
antihypernuclei yields in Au+Au collisions at RHIC using
the above mentioned parametrizations of thermal
parameters that best fit hadron
production at RHIC.
The present analysis uses a thermal model 
and  aims to
 elucidate the production mechanism of hypernuclei and
 antihypernuclei in heavy ion collisions at RHIC and LHC energies,
thus providing insight in the surprising increase of (anti)hypernuclei production
at high energies.
\section{\label{model}The THERMUS model}
The thermal model assumes that at freeze-out all hadrons in 
the hadron gas resulting from a high energy collision follow 
equilibrium distributions. The conditions at chemical freeze-out (when
inelastic collisions cease) are given by the hadron abundances, while
the particle spectra offer insight into the conditions at thermal
freeze-out (when elastic collisions cease). 
Once evaluated the hadron gas partition function gives all primordial 
thermodynamic quantities of the system by simple differentiation. The exact 
form of the partition function, however, depends on the statistical ensemble 
under consideration.\\
Within the grand-canonical ensemble the quantum numbers of the  
system are conserved on average through the action of chemical 
potentials~\cite{review}. 
In other words, the baryon content $B$, strangeness content $S$ 
and charge content $Q$ are fixed on average by $\mu_B$, $\mu_S$ 
and $\mu_Q$ respectively. For each of these chemical potentials 
one can write a corresponding fugacity using the standard prescription 
$\lambda=e^{\mu/T}$, where $T$ is the temperature of the system.\\
 
 As an example, the density of hadron species $i$ with 
quantum numbers $B_i$, $S_i$ and $Q_i$, spin-isospin degeneracy
factor, $g_i$, and mass, $m_i$,  
emitted directly from the fireball at temperature $T$ is given by
 a second order modified Bessel function of the second kind, 
\begin{equation}
\tilde{n_i}(T,\mu_B,\mu_S,\mu_Q,\gamma_S) =
\frac{g_i}{2\pi^2}m_i^2T\lambda_B^{B_i}\lambda_S^{S_i}\lambda_Q^{Q_i}\gamma_s^{|\tilde{S}_i|}K_2(\frac{m_i}{T}).   
\end{equation}
in the Boltzmann approximation.\\
The quantum-statistical result requires either an infinite summation over such 
$K_2$ functions or else a numerical integration~\cite{thermus1,thermus2}.\\

The chemical potentials $\mu_S$ and $\mu_Q$ are typically constrained
in applications of the model  
by the initial strangeness and baryon-to-charge ratio in the system under consideration.
\section{\label{antibaryon}Production of antibaryons}
\begin{figure}
\includegraphics[width=0.99\linewidth]{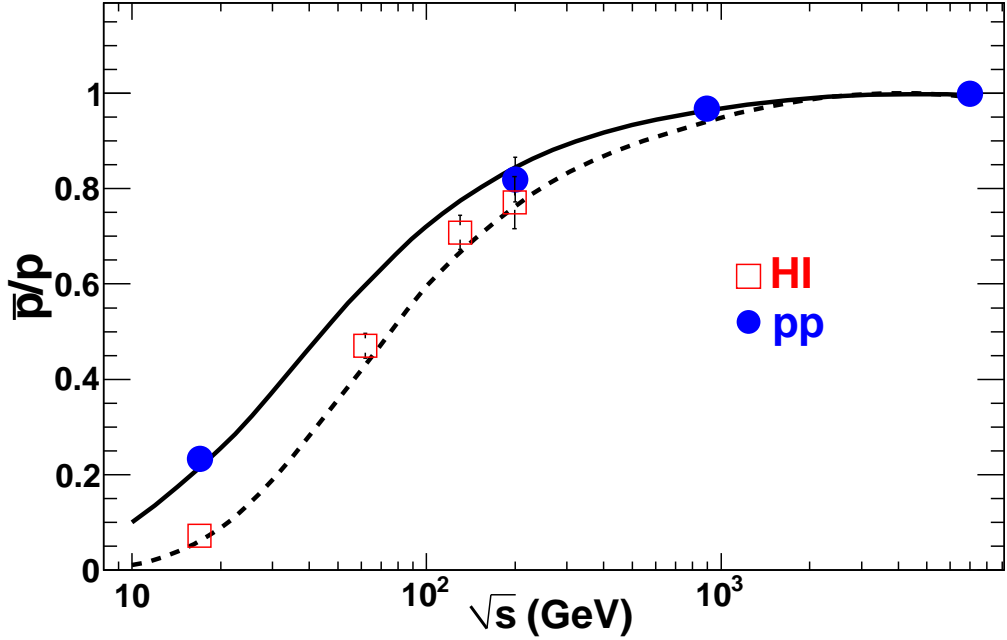}
\caption{The  \pbar/p ratio as  function of $\sqrt{s}$.
The solid circles are results from \pp collisions and the open squares
are results from HI collisions as a function
 of the invariant beam energy\cite{Alt:2005gr,Abelev:2006cs,Aamodt:2010d,Alt:2007fe,star:2009}.}
\label{pbarp}
\end{figure}

\begin{figure}
\includegraphics[width=0.99\linewidth]{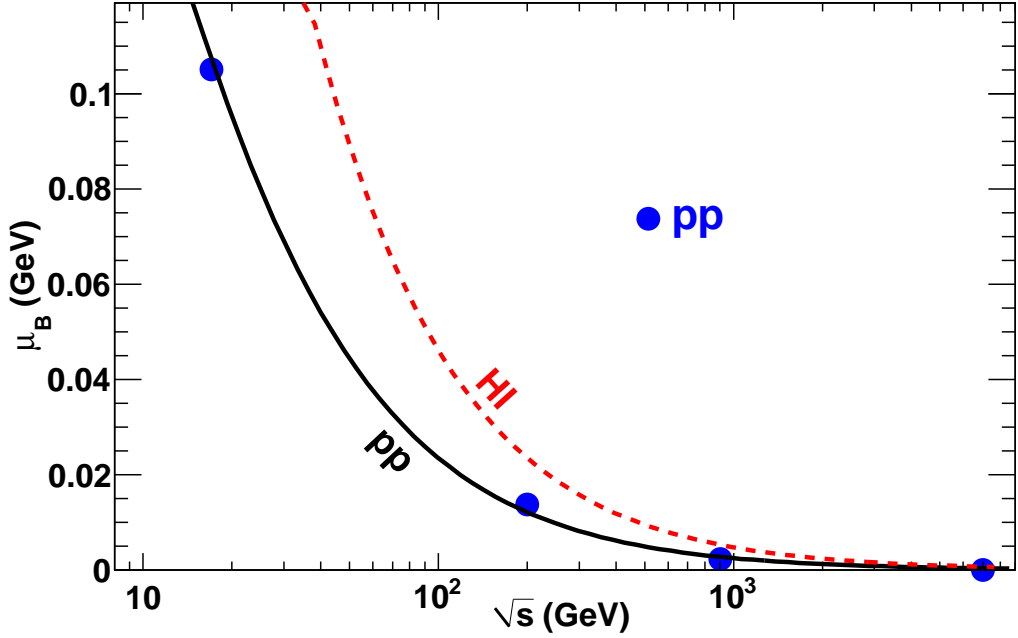}
\caption{Variation of the  baryon chemical potential {\muB} as a
function of $\sqrt{s}$. The dashed line describes heavy ion
collisions as in Ref.~\cite{Cleymans:2005xv} while the
 solid line is the new parametrization for pp collisions. } \label{MuBvsE}
\end{figure}

In heavy-ion collisions the increase in the  antimatter to matter ratio
with the center-of-mass energy of the system has been observed
earlier by the NA49~\cite{Alt:2005gr,Alt:2007fe} and the
STAR~\cite{Abelev:2006cs} collaborations.
The trend of \pbar/p ratio increase with the energy towards unity
is shown in Fig.~\ref{pbarp}, where the open squares refer
to heavy ion collisions and the solid circles refer to \pp collisions.
It include
results from the NA49~\cite{Alt:2005gr}, STAR~\cite{Abelev:2006cs} and the
new results from the ALICE Collaboration~\cite{Aamodt:2010d}.
The resulting baryon chemical potential $\mu_B$ is shown in  Fig.~\ref{MuBvsE} where
the dashed line refers to the heavy ion description using the THERMUS
model~\cite{thermus1,thermus2}.
The two input parameters, the chemical freeze-out temperature $T$ and
the baryon chemical potential {\muB} as a function of $\sqrt{s}$ are
taken from Ref.~\cite{Cleymans:2005xv}.
\begin{equation}
T(\mu_B) = a - b\mu_B^2 -c \mu_B^4  \label{Eqn:T(muB)}
\end{equation}
with $ a =  0.166 \pm 0.002$ GeV, $b = 0.139 \pm 0.016$
GeV$^{-1}$ and $c = 0.053 \pm 0.021$ GeV$^{-3}$. This
parametrization is similar and consistent with
the one proposed in Ref.~\cite{Andronic:2005yp}.
The solid line in Fig.~\ref{pbarp} is obtained from THERMUS 
model~\cite{thermus1,thermus2}
using $T$ from equation 1 and {\muB} from equation 2.
The solid circles represent {\muB}, obtained after fitting
experimental data with the THERMUS model~\cite{thermus1,thermus2}.
The solid line is a new
parametrization adjusted for pp collisions.
In view of the fact that peripheral and central collisions show
no noticeable change in the temperature we have used the
same $T$ dependence for \pp as in heavy ion collisions but the dependence
on
$\mu_B$ on beam energy is now given by
\begin{equation}
\mu_B = d / ( 1 + e \sqrt{s})
\end{equation}
with $d$ = 0.4 GeV and $e = 0.1599$ GeV$^{-1}$.

It is important to note that {\muB} is always lower
in pp collisions than in heavy
ion collisions,
e.g. the freeze-out chemical potential
follows a different pattern, due to the
lower stopping power in pp collisions.

The relation between the \pbar/p ratio and {\muB} can be shown
easily within the statistical concept using the Boltzmann
statistics Ref.~\cite{BraunMunzinger:2003zd}. In the model
calculation, the appropriate statistics
and also feed down from strong decays are taken into account.
The density of particle $i$ is then given by
\begin{equation}
n_i = \frac{d_i}{2\pi^2} \, K_2
\left(\frac{m_i}{T}\right) \, e^{(N_B \mu_B +N_S \mu_S)/T}  \label{eq:density}
\end{equation}
with $N_B$ and $N_S$ being the baryon and strangeness quantum numbers of
particle $i$.

This leads to a \pbar/p ratio of (excluding feed-down from heavier resonances):
\begin{equation}
\frac{n_{\rm\overline{p}}}{n_{\rm p}} =  e^{-(2 \mu_B)/T}
\end{equation}

The ratio of strange antibaryons/ baryons is then given by
\begin{equation}
\frac {n_{\rm\overline{B}}}{n_{\rm B}} =  e^{-(2 \mu_B - N_S \mu_S
)/T}
\end{equation}

\begin{figure}
\includegraphics[width=0.99\linewidth]{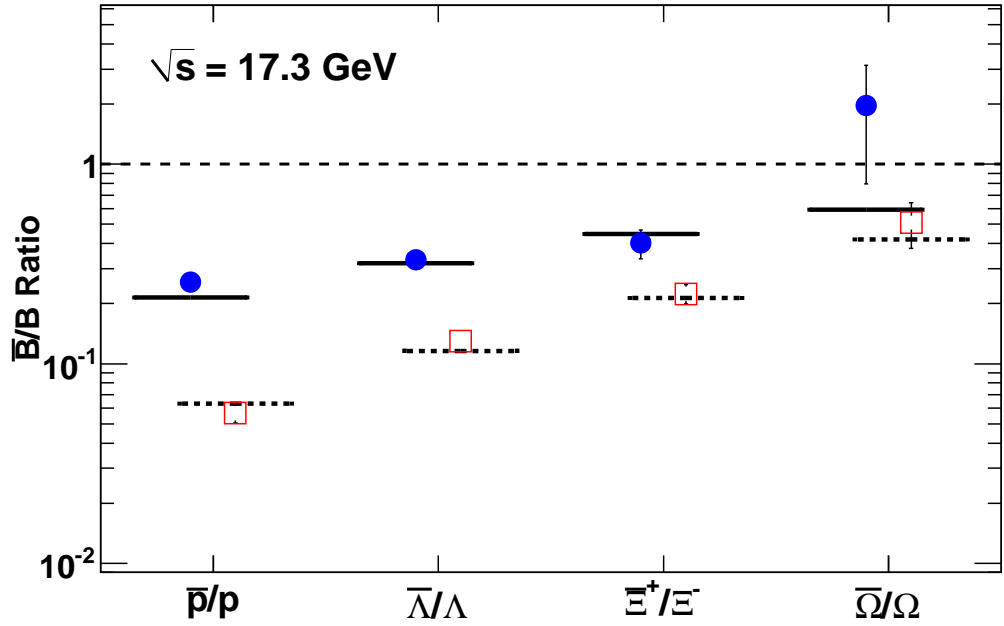}
\caption{
Antibaryon to baryon ratios at the SPS according to
strangeness content.
Circles refer to \pp collisions, squares to heavy ion collisions.
\label{ratio_SPS} }
\end{figure}

\begin{figure}
\includegraphics[width=0.99\linewidth]{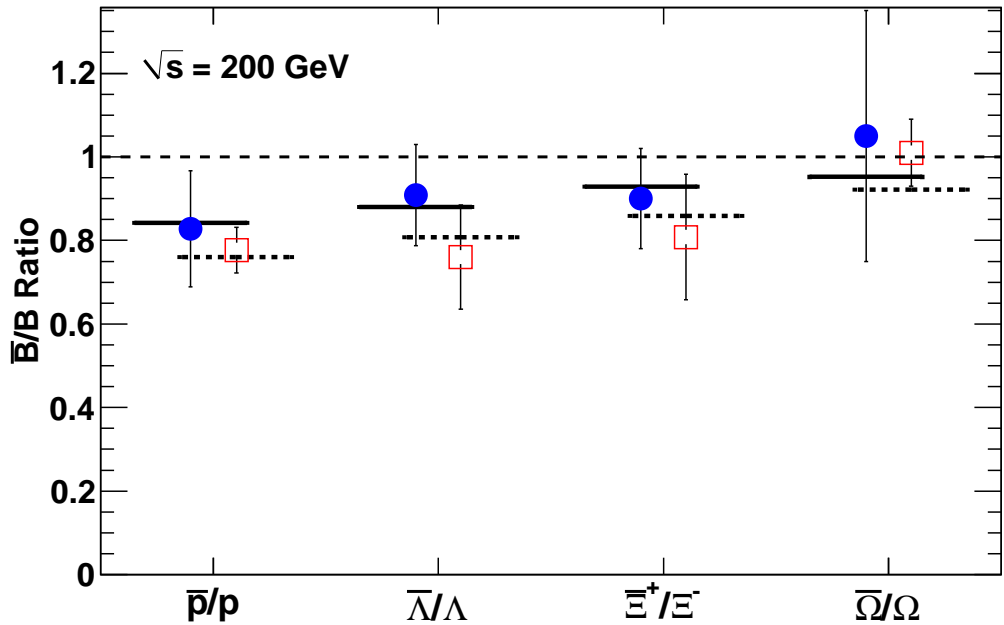}
\caption{
Antibaryon to baryon ratios at STAR according to
strangeness content.
Circles refer to \pp collisions, squares to heavy ion collisions.
\label{ratio_RHIC} }
\end{figure}

As \muS~ is always smaller than \muB , the ratios appear ordered
with the strangeness quantum number, i.e.~the higher $N_S$, the
smaller the difference between antibaryon and baryon. This trend
is shown in Figs.~\ref{ratio_SPS} and \ref{ratio_RHIC} comparing
the results from the model with experimental data.
The agreement between the model results and the data is very good.

\section{\label{secAnalysis} Production of nuclei, antinuclei,  hypernuclei and
antihypernuclei}

\subsection{Comparison to data from RHIC}

The production of light nuclei including hypertritons
 ($^3_\Lambda$H) and antihypertritons ($^3_{\bar{\Lambda}}\overline{\textrm{H}}$)
was recently observed by the STAR collaboration~\cite{Abelev:2010}.
The abundances of such light nuclei and antinuclei follows a 
consistent pattern in the thermal model. 
The temperature remains the same as before but an extra 
factor of $\mu_B$ is picked up each time the baryon number is increased.
Each proton or neutron thus simply adds a factor of $\mu_B$ to the 
Boltzmann factor. 
The production of nuclear fragments is therefore very sensitive to the 
precise value of the baryon chemical potential and could thus lead
to a precise determination of $\mu_B$.

The ratios within the statistical approach using the
grand-canonical formalism can be easily written, based on
Eq.~(\ref{eq:density}).
Deuterium has an additional neutron and the antideuterium to 
deuterium ratio is given by the square of the antiproton to 
prton ratio:
\begin{equation}
\frac {n_{\rm\overline{d}}}{n_{\rm d}}       =  e^{-(4 \mu_B )/T}
\end{equation}

Helium 3 has 3 nucleons and the corresponding anti-Helium 3 to
helium 3 ratio is given by: 
\begin{equation}
\frac {n_{\rm ^3\overline{He}}}{n_{\rm ^3He}} =  e^{-(6 \mu_B)/T}
\end{equation}
If the nucleus carries strangeness this leads to an extra factor of $\mu_S$
\begin{equation}
\frac {n_{\rm ^3_{\overline{\Lambda}}\overline{H}}}{n_{\rm ^3_{\Lambda}H}} =
e^{-(6 \mu_B - 2 \mu_S )/T}
\end{equation}
In mixed ratios  the different degeneracy factors are also taken
into account, e.g. 6 for $^3_\Lambda H$ 
and 2 for $^3_\Lambda H$ .
\begin{equation}
\frac {n_{\rm ^3_{\Lambda}H}}{n_{\rm ^3He}}  = 3e^{-(6 \mu_B
- \mu_S )/T}
\end{equation}
In the model like in the data the $He^3$ and $\overline{He^3}$ yields 
have been corrected
for the part coming from hypertriton and antihypertriton decays
assuming a decay branch ratio for the decay 
of 25 \%.
\subsection{Predictions for RHIC and LHC}
\begin{figure}
\includegraphics[width=0.99\linewidth]{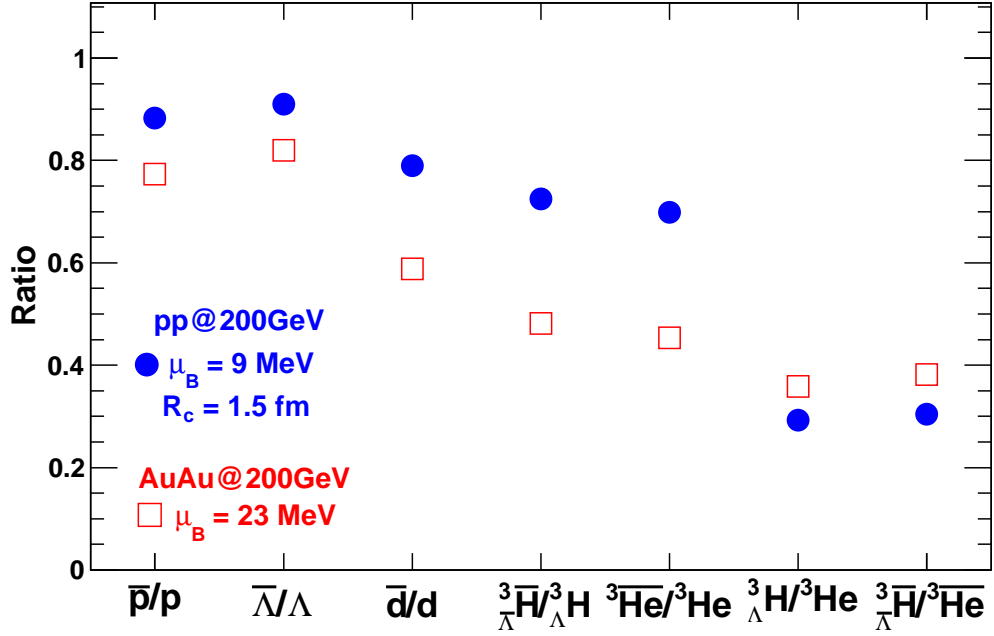}
\caption{Comparison of pp and heavy ion collisions at \s = 200 GeV
evidencing the influence of different values of $\mu_B$ and of the
canonical suppression.}
 \label{AuAu_pp_200}
\end{figure}

In Fig.~\ref{AuAu_pp_200} we compare \pp and heavy ion collisions 
at $\sqrt{s} = 200$ GeV. The difference between the 
two colliding systems and the effect of 
canonical suppression is seen in \pp collisions.

\begin{figure}
\includegraphics[width=0.99\linewidth]{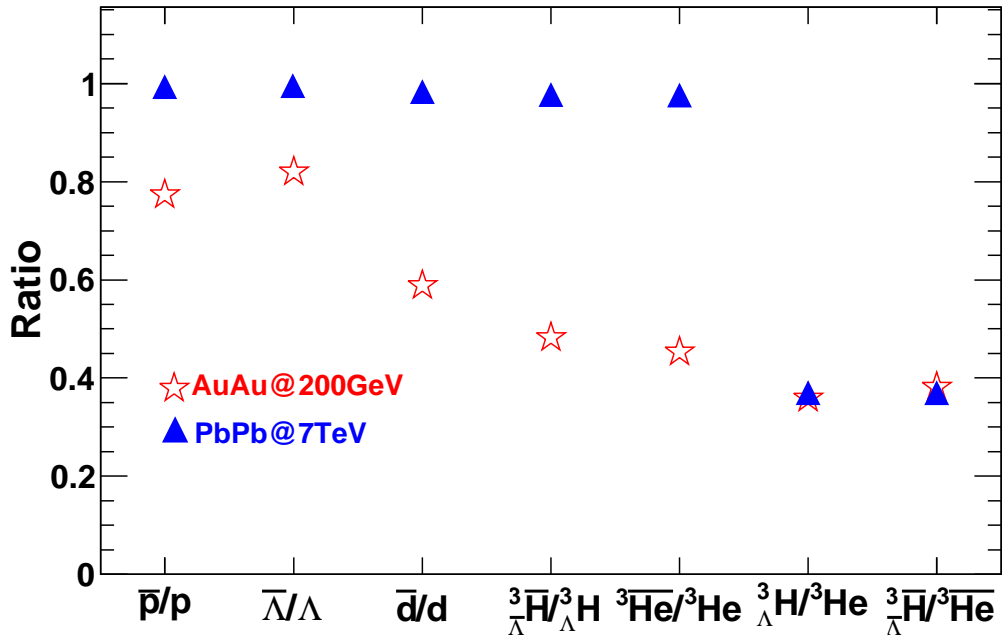}
\caption{Comparison of two different collision energies for heavy
ion collisions} \label{comp_200_7}
\end{figure}

In Fig. 7 a comparison is shown of the various  antiparticle/partcle
ratios for two different beam energies.

\begin{figure}
\includegraphics[width=0.99\linewidth]{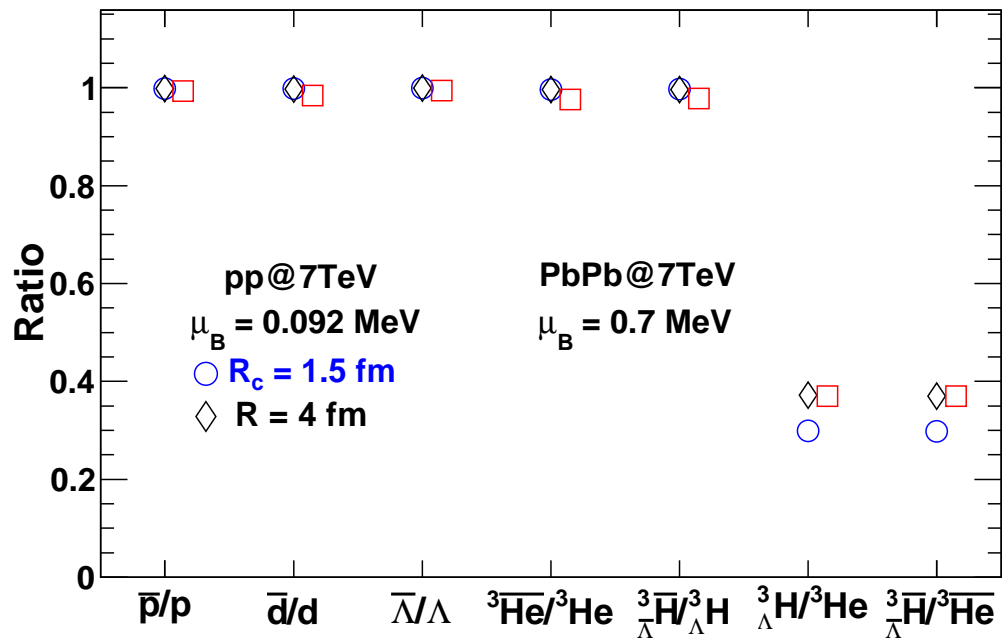}
\caption{Prediction for \s\ = 7 TeV both for pp and PbPb collision.}
\label{pred_7}
\end{figure}

The expectations for the LHC are shown in Fig.~\ref{pred_7}.

\begin{figure}[hbt]
\includegraphics[width=0.99\linewidth]{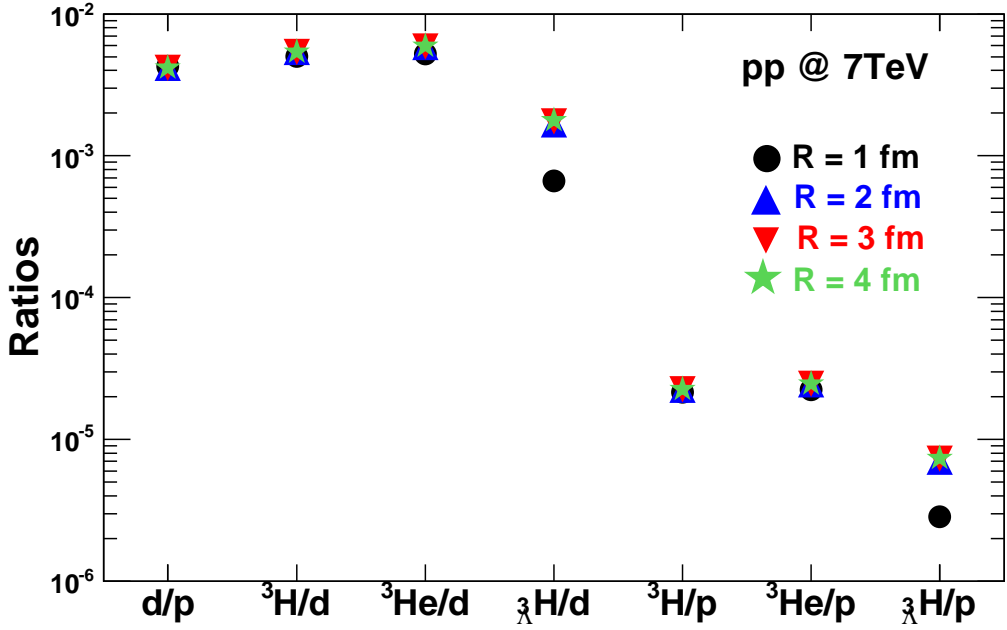}
\caption{The ratio of the yield for examples of different masses.}
\label{mass_diff}
\end{figure}

Finally the predictions of the thermal model for
ratios of anti-nuclear to nuclear fragments are shown in 
Fig.~\ref{mass_diff}.
This figure includes comparisons for strange nuclear fragments 
where a clear picture emerges (again) between strange and non-strange 
fragments.


\section{\label{secSummary} Discussion and Summary}
In the present paper we have made a general comparison of
thermal parameters in \pp and heavy ion collisions. We
have determined the energy dependence of the baryon chemical potential
\muB  in \pp collisions.  This was used to establish a hierarchy
of antibaryon to baryon ratios including strange and multi-strange baryons.
This was then used to compare nuclear and anti-nuclear fragments in \pp
and heavy ion collisions.
Predictions have been presented for these ratios at LHC energies.
\begin{acknowledgments} 
We acknowledge the support of DFG, the Polish  Ministry of Science MEN, the
CNRS, IN2P3 (France) 
and the Alexander von Humboldt Foundation. The financial support of the BMBF,
the DFG-NRF, the Department of Science and Technology of the Government of India and the South Africa - Poland scientific collaborations are also gratefully acknowledged.
\end{acknowledgments}
%


\end{document}